\documentclass[epj]{svjour}
\usepackage{graphicx}
\usepackage{amssymb}
\newcommand{\bea}{\begin{eqnarray}}
\newcommand{\eea}{\end{eqnarray}}
\newcommand{\nn}{\nonumber}
\newcommand{\nnl}{\nonumber\\}
\begin{document}
\title{Spectral Functions of Quarks in Quark Matter\thanks{Work supported
by DFG.}}
\author{F. Fr\"omel \and S. Leupold \and U. Mosel}
\institute{Institut f\"ur Theoretische Physik, Universit\"at Giessen,
D-35392 Giessen, Germany}
\date{Received: date / Revised version: date}
\abstract{
We present a simple albeit self-consistent approach to the spectral function of
light quarks in infinite quark matter. Relations between correlation functions
and collision rates are used to calculate the spectral function in an iterative
procedure. The quark interactions are described by the SU(2) Nambu--Jona-Lasinio
model. Calculations were performed in the chirally restored phase at zero
temperature.
} 
\maketitle


\section{Introduction}
\label{sec:intro}

It is well known that short-range correlations have influence on the properties
of nuclear matter and finite nuclei; a substantial amount of high-momentum
processes is contained in the nucleon spectral function. There have been many
theoretical approaches trying to understand the short-range correlations. Of
particular interest are the self-consistent calculations for the spectral
function of nucleons in nuclear matter from Lehr et al. \cite{le}. The results
of their simple model are in good agreement with sophisticated calculations
from many-body theory. It is striking that their model is very successful in
describing the influence of short-range correlations on the properties of
nuclear matter using a simple pointlike nucleon interaction with
a constant scattering amplitude.

Motivated by the success of this model we have taken up the concept to
investigate the properties of light quarks in infinite quark matter \cite{ff}.
It is our basic assumption that the properties of the spectral function are
dominated by phase space effects and the overall strength of the interaction
just like in the case of the nucleons. The detailed structure of the
interaction should be relatively unimportant as long as the relative symmetries
are respected. We use relations between the spectral function and the
collisional self-energies to construct a simple albeit self-consistent model.
The spectral function can then be calculated in an iterative process beyond the
quasiparticle approximation. The quark interactions are described by the
Nambu--Jona-Lasinio (NJL) model \cite{klev}. It has the same symmetries as QCD
and describes an effective pointlike interaction.

As a first step the model has been applied to the simplest system, namely the
chirally restored phase at zero temperature. Mean field effects were neglected,
diquark condensates that arise in the color superconducting phase were not
included. For the collisional self-energies only the lowest order
contributions, the Born diagrams, were considered. In the numerical
calculations the influence of the coupling strength and the chemical potential
on the properties of the spectral function was investigated.

\section{The Model}
\label{sec:model}

In this section we will briefly review our model. For more details we refer to
\cite{ff}. The underlying Green's function formalism is discussed in much
detail in \cite{kb,dan}. Note that current quark masses are neglected
throughout this work and only systems in thermal equilibrium are considered. We
use the Nambu--Jona-Lasinio model to describe the quark interactions in our
approach. It is an effective interaction model that was designed to resemble
the symmetries of QCD. The SU(2) Lagrangian is given by
\bea
        \mathcal{L}_\mathrm{NJL}= \bar \psi i \partial\!\!\!/ \psi
        + G [(\bar \psi \psi)^2+(\bar \psi  i \gamma_5 \vec \tau \psi)^2],
\eea
where $G$ is the coupling strength, independent of energy and momentum, and the
$\tau_i$ are isospin Pauli matrices. Due to the pointlike interaction this
model is nonrenormalizable and a three momentum cutoff $\Lambda$ is introduced.
Currently we do not consider any extensions to the Lagrangian that lead to
color superconductivity, cf. \cite{ff}.

The correlation functions
\bea
        i g^>(1,1') &=& \langle\psi(1) \bar\psi(1') \rangle, \nnl
 -i g^<(1,1') &=& \langle \bar\psi(1') \psi(1) \rangle \nn
\eea
are the fundamental elements of our model. In thermal equilibrium they are
related to the spectral function $\mathcal{A}$ via the thermal Fermi
distribution $n_F(p_0)$:
\bea
        -ig^<(p)&=&\mathcal{A}(p)n_F(p_0),      \label{eq:gltspec} \\
        ig^>(p)&=&\mathcal{A}(p)[1-n_F(p_0)].   \label{eq:ggrspec}
\eea
Note that the Green's functions and the spectral function are matrices in
spinor space. The single particle self-energy can be decomposed into a
mean-field part $\Sigma^\mathrm{mf}$ and the collisional self-energies
$\Sigma^\gtrless$. Thus the retarded self-energy $\Sigma^\mathrm{ret}$ is given
by:
\bea
        \Sigma^\mathrm{ret}(1,1') &=&  \Sigma^\mathrm{mf}(1,1') \nnl
        &+& \Theta(t_{1}-t_{1'})\left[\Sigma^>(1,1')-\Sigma^<(1,1')\right].
        \label{eq:sigmaret}
\eea

\begin{figure}
        \centering{\includegraphics[scale=.55]{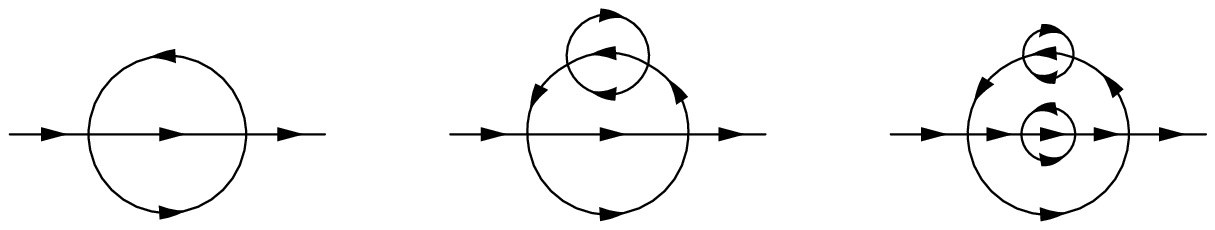}}
        \caption{\label{fig:selfconsist}Examples for diagrams that are resummed
        in our calculation. The lines correspond to free propagators.}
\end{figure}

The time local mean-field self-energy corresponds to the motion of particles in
a potential. It is responsible for the dynamical generation of the constituent
quark masses. The collisional self-energies contain the effects of particle
decays and collisions in the medium. In lowest order they are given by the Born
diagrams (cf. left diagram in Fig. \ref{fig:selfconsist}). One finds
\bea
        \pm i\Sigma^\gtrless(p) &\sim& \int\!\!\!\int\!\!\!\int \cdots
        G^2 g^\lessgtr(p_2)g^\gtrless(p_3)g^\gtrless(p_4), \label{eq:sigma}
\eea
where $g^>$ and $g^<$ are full propagators and the integrals run over
$p_2,p_2,$ and $p_4$. Eqs. (\ref{eq:sigma}) look like total collision rates and
can be used to determine the width of the spectral function:
\bea
        \Gamma(p)=-2\mathrm{Im}\Sigma^\mathrm{ret}(p)
        =i[\Sigma^>(p)-\Sigma^<(p)]. \label{eq:gamma}
\eea
The real part of $\Sigma^\mathrm{ret}$ is related to $\Gamma$ by a dispersion
relation. If the width is known over the full energy range $\mathrm{Re}
\Sigma^\mathrm{ret}$ can be calculated dispersively. Using $\Gamma$ and
$\mathrm{Re}\Sigma^\mathrm{ret}$ the spectral function can be explicitly
written as:
\bea
        \mathcal{A}(p) = -2\mathrm{Im}g^\mathrm{ret}(p)
                             = -2\mathrm{Im}\frac{1}{\not\!p
                                 -\mathrm{Re}\Sigma^\mathrm{ret}+i\Gamma/2}.
        \label{eq:spec}
\eea
Eqs. (\ref{eq:gltspec}),(\ref{eq:ggrspec}), and
(\ref{eq:sigma})-(\ref{eq:spec}) form a set of equations describing a
self-consistency problem. A direct solution is not easily possible. It is
possible, however, to solve the problem iteratively by starting from an initial
guess for one of the quantities. In this way we can find a solution for the
spectral function. The effect of self-consistency can be nicely illustrated in
the language of Feynman diagrams. Using full propagators $g^\gtrless$ that
depend themselves on $\Sigma^\gtrless$ in eqs. (\ref{eq:sigma}) means that we
sum nonperturbatively over a whole class of diagrams. Some examples are shown
in Fig. \ref{fig:selfconsist}.

Finally, we have to discuss the matrix structure of the spectral function in
spinor space. To find the most general form $\mathcal{A}$ can be decomposed in
terms of the 16 independent products of the $\gamma$ matrices. Demanding
invariance under parity and time reversal symmetry we find for the rest frame
of the medium (in thermal equilibrium) \cite{bd}:
\bea
        \mathcal{A}(p)=\rho_\mathrm{s}(p_0,\vec p^{\,2})
        +\rho_\mathrm{0}(p_0,\vec p^{\,2})\gamma^0
        +\rho_\mathrm{v}(p_0,\vec p^{\,2})\hat{\vec p}\cdot \vec \gamma,
        \label{eq:structure}
\eea
where $\hat{\vec p}$ is a unit vector in the momentum direction. Note that the
density of states is given by $\rho_0$ alone and that $\rho_\mathrm{s}$ has to
be zero in the chirally restored phase. It follows from eqs.
(\ref{eq:gltspec})-(\ref{eq:spec}) that $\Gamma$ and $\Sigma^\gtrless$ must
have the same structure as $\rho$.

Currently we apply two simplifications to the model. First, we restore chiral
symmetry 'manually' by setting $\Sigma^\mathrm{mf}$ to zero. This can be done
for any density and temperature since $m=0$ is always a solution of the
equation for the constituent quark mass \cite{klev}. However, one has to be
aware that this might not be the thermodynamically favored phase when also a
finite solution for $m$ exists. Second, we neglect the real part of
$\Sigma^\mathrm{ret}$ due to technical reasons \cite{ff}. For nuclear matter
this was a reasonable approximation \cite{le}. To make sure that the effects of
this violation of analyticity are not significant we have checked that
$\mathrm{Re}\Sigma^\mathrm{ret}$ is small compared to $\not\!p$, cf.
eq.(\ref{eq:spec}).

\section{Results}
\label{sec:results}

\begin{figure}
        \centering{\includegraphics[scale=.43]{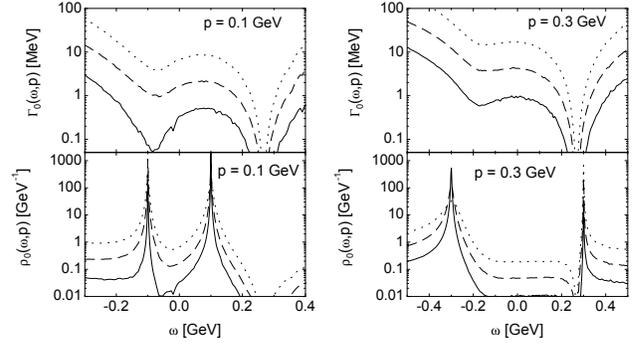}}
        \caption{\label{fig:coupling}The width $\Gamma_0$ and the spectral
        function $\rho_0$ of quarks at different momenta. Solid lines
        correspond to the usual NJL coupling strength, dashed lines to a
        coupling twice as large. Dotted lines have been obtained with a
        coupling four times larger than the usual value.}
\end{figure}

All our calculations were performed at zero temperature and in the chirally
restored phase. The calculations were initialized with a constant width,
$\Gamma_0=1\,\mathrm{MeV}$ and $\Gamma_\mathrm{v}=0$. Self-consistency  was
achieved after two iterations. First, we chose the quark matter density such
that it is comparable to regular nuclear matter, $\rho_\mathrm{qm}
=3\cdot\rho_\mathrm{nm}=3\cdot 0.17 \,\mathrm{fm}^{-3}$. This yields a Fermi
energy of $\omega_F=0.268\,\mathrm{GeV}$. The cutoff $\Lambda$ and the coupling
constant $G$ of the NJL model were chosen so that the model reproduces the
known values \cite{klev} for the quark condensate and the pion coupling
constant $f_\pi$ in vacuum. To investigate the influence of the coupling on the
spectral function we did also calculations with two times and four times larger
coupling strengths. In reality quark matter with a chemical potential of
$\omega_F=0.268\,\mathrm{GeV}$ would not be in the chirally restored phase.
Therefore we made additional calculations with a higher density of
$\rho_\mathrm{qm}=1.53 \,\mathrm{fm}^{-3}$, corresponding to
$\omega_F=0.387\,\mathrm{GeV}$. This chemical potential is well beyond the
chiral phase transition in the NJL model \cite{klev}.

Fig. \ref{fig:coupling} shows our results for the spectral function and its
width at several momenta and coupling strengths using $\omega_F=
0.268\,\mathrm{GeV}$. Due to the pointlike interaction the width seems to
increase explosively for large $|p_0|$. At even higher $|p_0|$, however,
the opening of phase space is suppressed by the NJL cutoff and the width
decreases again. Physically the most interesting area lies in the energy range
$0<p_0<\omega_F$ since that is the region of the populated quark states. All
states above the Fermi energy as well as the anti-quark states at negative
$p_0$ are unoccupied (no holes in the Dirac sea). The structure of the spectral
function is dominated by the on-shell peaks of the quarks and anti-quarks at
$p_0=|\vec p|$ and $p_0=-|\vec p|$. The peaks get broader when the coupling is
increased. Strength is distributed away from the peaks to the off-shell
regions, the width of the peaks increases from $0.1-1\,\mathrm{MeV}$ to
$10\,\mathrm{MeV}$. The width seems to scale with the coupling strength (resp.
$G^2$, cf. eqs. (\ref{eq:sigma})) while the general shape remains unchanged.

\begin{figure}
        \centering{\includegraphics[scale=.43]{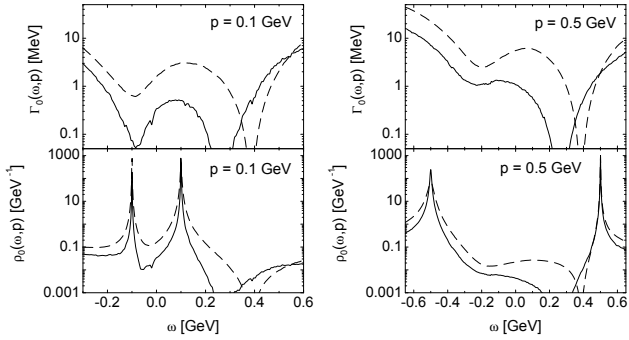}}
        \caption{\label{fig:chempot}The width $\Gamma_0$ and the spectral
        function $\rho_0$ of quarks at different momenta. Solid lines
        correspond to $\omega_F=0.268\,\mathrm{GeV}$, dashed lines to
        $\omega_F=0.387\,\mathrm{GeV}$.}
\end{figure}

Fig. \ref{fig:chempot} shows the width and the spectral function for the two
chemical potentials $\omega_F=0.268\,\mathrm{GeV}$ and $\omega_F=
0.387\,\mathrm{GeV}$ using the regular coupling strength. The effect of the
higher density is comparable to increasing the coupling. The width
approximately scales with the chemical potential while the shape remains
unchanged. This leads again to a broadening of the peaks of the spectral
function.

\begin{figure}
        \centering{\includegraphics[scale=.43]{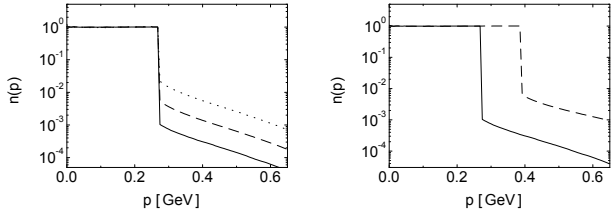}}
        \caption{\label{fig:momdist} \textbf{(a)} Quark momentum distribution
        in quark matter for $\omega_F=0.268\,\mathrm{GeV}$ and three different
        couplings (see Fig. \ref{fig:chempot} for details). \textbf{(b)}
        Momentum distribution for two different chemical potentials (see Fig.
        \ref{fig:coupling} for details).}
\end{figure}

In the momentum distribution of nucleons in nuclear matter a depletion of the
occupation probabilites by about 10\% is seen \cite{le}. The resulting high
energy tail is taken as a universal sign of short-range correlations. We show
the momentum distribution of the quarks for the different coupling strengths
in Fig. \ref{fig:momdist}(a). At the lowest coupling a depletion of only 0.1\%
is found. For the coupling twice as large the short-range correlations increase
but still the depletion effect is below 1\%. Only for the largest couplings we
find a high momentum tail of a few percent, comparable to the case of nucleons.
In Fig. \ref{fig:momdist}(b) it can be seen again that the effect of the larger
chemical potential is similar to an increased coupling. The depletion for
$\omega_F=0.387\,\mathrm{GeV}$ grows by almost one order of magnitude compared
to $\omega_F=0.268\,\mathrm{GeV}$.

\section{Summary and Outlook}
\label{sec:outlook}

Based on a successful model for nuclear matter we have presented a simple but
fully self-consistent approach to the spectral function of quarks in quark
matter. It uses a pointlike interaction and goes beyond the quasiparticle
approximation. The numerical results indicate that the influence of short-range
correlations is rather small compared to nuclear matter. This finding might be
an artifact of the present model, the NJL model with vacuum par\-a\-me\-ters in 
the Born approximation. Furthermore we have not considered broken symmetries.
However, the calculations have shown that the model is technically feasible and
suitable for further development.

\begin{figure}
        \centering{\includegraphics[width=\columnwidth]{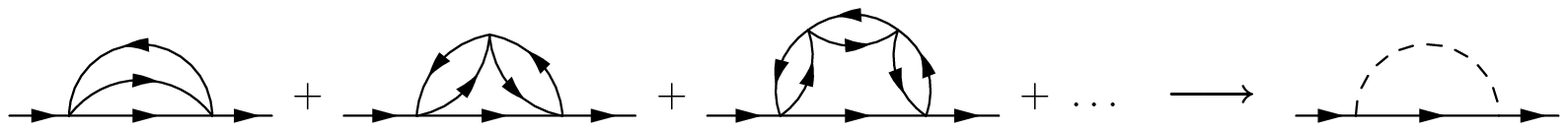}}
        \caption{\label{fig:dynmesons}Series of diagrams corresponding to
        dynamically generated mesons.}
\end{figure}

At present, the model is extended in two ways. First, a more realistic phase
with broken symmetries -- the chirally broken phase -- is considered. In this
phase large constituent quark masses must be taken into account and the
analytic structure of the spectral function (\ref{eq:structure}) is more
complicated. Second, a new class of diagrams is incorporated into the
collisional self-energies (\ref{eq:sigma}). This series of diagrams is shown in
Fig. \ref{fig:dynmesons}. The lowest order contribution is again the Born
diagram. The full series can be interpreted as a dynamically generated meson
that couples to the quarks \cite{klev,reh}. First estimates show that this
extension will increase the effective quark coupling and should lead to
significantly higher widths of the spectral function. In addition it will be
possible to investigate the properties of the self-consistently calculated
pions and sigmas.

\end{document}